\documentclass[prl,aps,amsmath,amssymb,twocolumn,groupedaddress,floats,showpacs,final]{revtex4-1}
\usepackage{graphicx}
\usepackage{subfigure}
\usepackage{dcolumn}
\usepackage{bm}
\usepackage{color}
\usepackage{hyperref}
\begin{document}
\author {Kun Chen$^{1,2}$,Longxiang Liu$^{1}$}
\author {Youjin Deng$^{1,2}$}
\email{yjdeng@ustc.edu.cn}
\author {Lode Pollet$^{3}$}
\email{lode.pollet@physik.uni-muenchen.de}
\author {Nikolay Prokof'ev$^{2,4}$ }
\email{prokofev@physics.umass.edu}
\affiliation{$^{1}$ National Laboratory for Physical Sciences at Microscale and Department of Modern Physics, University of Science and Technology of China, Hefei, Anhui 230026, China}
\affiliation{$^{2}$Department of Physics, University of Massachusetts, Amherst, Massachusetts 01003, USA}
\affiliation{$^{3}$Department of Physics and Arnold Sommerfeld Center for Theoretical Physics, Ludwig-Maximilians-Universit{\"a}t M{\"u}nchen, D-80333 M{\"u}nchen, Germany}
\affiliation{$^{4}$Russian Research Center ``Kurchatov Institute", 123182 Moscow, Russia }

\title{Universal Conductivity in a Two-Dimensional Superfluid-to-Insulator Quantum Critical System}

\date{\today}

\begin{abstract}
We compute the universal conductivity of the (2+1)-dimensional XY universality class, which is realized for a superfluid-to-Mott insulator quantum phase transition at constant density.
Based on large-scale Monte Carlo simulations of the classical (2+1)-dimensional $J$-current model and the two-dimensional Bose-Hubbard model, we can precisely determine the conductivity on the quantum critical plateau, $\sigma(\infty)=0.359(4)\sigma_Q$ with $\sigma_Q$ the conductivity quantum.
The universal conductivity curve is the standard example with the lowest number of components where the bottoms-up AdS/CFT correspondence from string theory can be tested and made to use [R. C. Myers, S. Sachdev, and A. Singh, Phys. Rev. D83,
066017 (2011)].
For the first time, the shape of the $\sigma(i\omega_n)- \sigma(\infty)$ function in the Matsubara representation is accurate enough for a conclusive comparison and establishes the particle-like nature of charge transport.
We find that the holographic gauge/gravity duality theory for transport properties
can be made compatible with the data if
temperature of the horizon of the black brane is different from the temperature of the conformal field theory.
The requirements for measuring the universal conductivity in a cold gas experiment are also determined by our calculation.

\end{abstract}
\maketitle
Transport properties of systems in the quantum critical region near zero-temperature have been a subject
of intense theoretical and experimental studies for decades. Systems in two spatial dimensions (2D)  fall in a particularly challenging class: Due to strong fluctuations standard mean field and perturbative methods fail, rendering the problem notoriously difficult and controversial.  The optical conductivity, $\sigma(\omega)$, is the key transport property describing the response of  particles to an externally applied frequency dependent chemical potential gradient, or ``electric'' field \cite{book}. Quantum critical points (QCP) in two-dimensional models with XY symmetry have
emergent symmetry larger than Lorentz symmetry, conformal invariance, described by a conformal field theory (CFT).
The conductivity then has zero scaling dimension and follows the scaling law~\cite{sachdev97},
\begin{equation}
\label{eq:scaling}
\sigma(\omega, T \rightarrow 0)=\sigma_Q \Sigma(\omega/T), \;\;\mbox{with} \;\;
\sigma(\omega/T \to \infty ) \to \sigma (\infty) \;,
\end{equation}
where we set $\hbar=k_B=1$ as units. The conductivity quantum, $\sigma_Q=2\pi Q^2$ (see Ref.~\cite{book}),
absorbs the coupling constant to the external field used to induce the gradient in the chemical potential, {\it e.g.},  the
charge of carriers $Q$ leaving $\Sigma(x)$ a dimensionless, universal scaling function.

Theoretically, a perturbative approach based on the leading order term in both the $\epsilon=3-d$ and $1/N$ expansions
has limited power to predict the quantitative and even qualitative properties of $\Sigma(x)$ in
$d=2$ and $N=2$\cite{book,fisher91,sachdev97,Fazio,william12}.
This impasse was broken by the promising introduction of the AdS/CFT or holographic correspondence from string theory to condensed matter physics\cite{mald,review1,review2}. The main results of the correspondence of interest to this Letter are that the strongly coupled CFT in the quantum critical region can be mapped onto a weakly coupled gravity theory in anti-de Sitter space (AdS), where $\Sigma(x)$ can be computed by standard perturbative techniques\cite{sachdev11,Subir13}.
When holographic theory is approximated by a classical gravity theory (and truncated to terms up to four derivatives)
the final result for $\Sigma(x)$ has only two free parameters: $\Sigma(\infty)$ and
$\Sigma(0)$, or $\gamma=(\Sigma(0)-\Sigma(\infty))/4\Sigma(\infty)$, meaning that the shape of the universal
$\Sigma(x)-\Sigma(\infty)$ curve is completely determined by $\gamma$.
The charge transport could either be particle-like for $\gamma>0$ or vortex-like for $\gamma<0$, where causality further constrains the value of $\gamma$ to be $| \gamma | \le 1/12$\cite{sachdev11,review1}.

It is currently not possible to carefully test the holographic theory neither in experiments nor in simulations.
On the one hand, optical conductivity measurements at frequencies $\omega/T \sim 1$ in 2D have so far been rather limited and ambiguous\cite{engel,crane}. On the other hand, previous \cite{sorensen05} as well as the most recent \cite{sorensen13} numerical studies were severely affected by the fact that simulations were performed away from the critical point on the insulating side of the transition, see the supplementary material \cite{suppl}.
This results in a systematic error big enough to distort the $\sigma (i\omega_n/T) - \sigma (\infty)$ function
to such a degree that the answer is no longer faithfully representing the universal curve.
For this reason our numerics are incompatible with Ref.~\cite{sorensen13} within the claimed error bars.

In this Letter, we obtain an accurate estimate of the $\sigma(\infty )$ parameter which is
(i) the key dynamic characteristic of the quantum critical continuum, and (ii) the most robust property of the quantum critical point, {\it e.g.} it is stable against disorder \cite{batr}
and detuning from the critical point \cite{gazit}.
We found that it was necessary to simulate system sizes that are at least an order of magnitude larger than in previous studies, which in turn required that the location of the QCP be refined to six significant digits. Only then were we able to deduce the universal function $\sigma(i\omega_n)-\sigma(\infty )$ with error bars suitable for precise tests of analytic theories, in particular the holographic theory\cite{sachdev11}.
We find that the holographic correspondence with temperature of the black brane horizon, $T_B$, equal to $T$ \cite{sachdev11,Subir13} cannot account for the data. 
The fit works better if one includes $T_B$ to the set of fitting parameters (an idea proposed in Ref.~\cite{sorensen13}) suggesting that the theory can be renormalized.
To go from the Matsubara to the real-frequency axis we also try to fit the data by a 
simple analytical form and obtain results consistent with the particle-like transport (as opposed to vortex-like transport).
By simulating the quantum critical liquid state of the Bose-Hubbard model, we shed light on the possibility
of measuring the universal conductivity with ultracold atoms in optical lattices. Given the extreme
sensitivity of the results on system parameters and the need of extrapolation from available system sizes,
one would expect that such an experiment will be challenging without theoretical support. However,
we demonstrate that $\sigma (\infty)$ may well be within reach.

\begin{figure}[htbp]
\includegraphics[scale=0.5,angle=0,width=1.0\columnwidth]{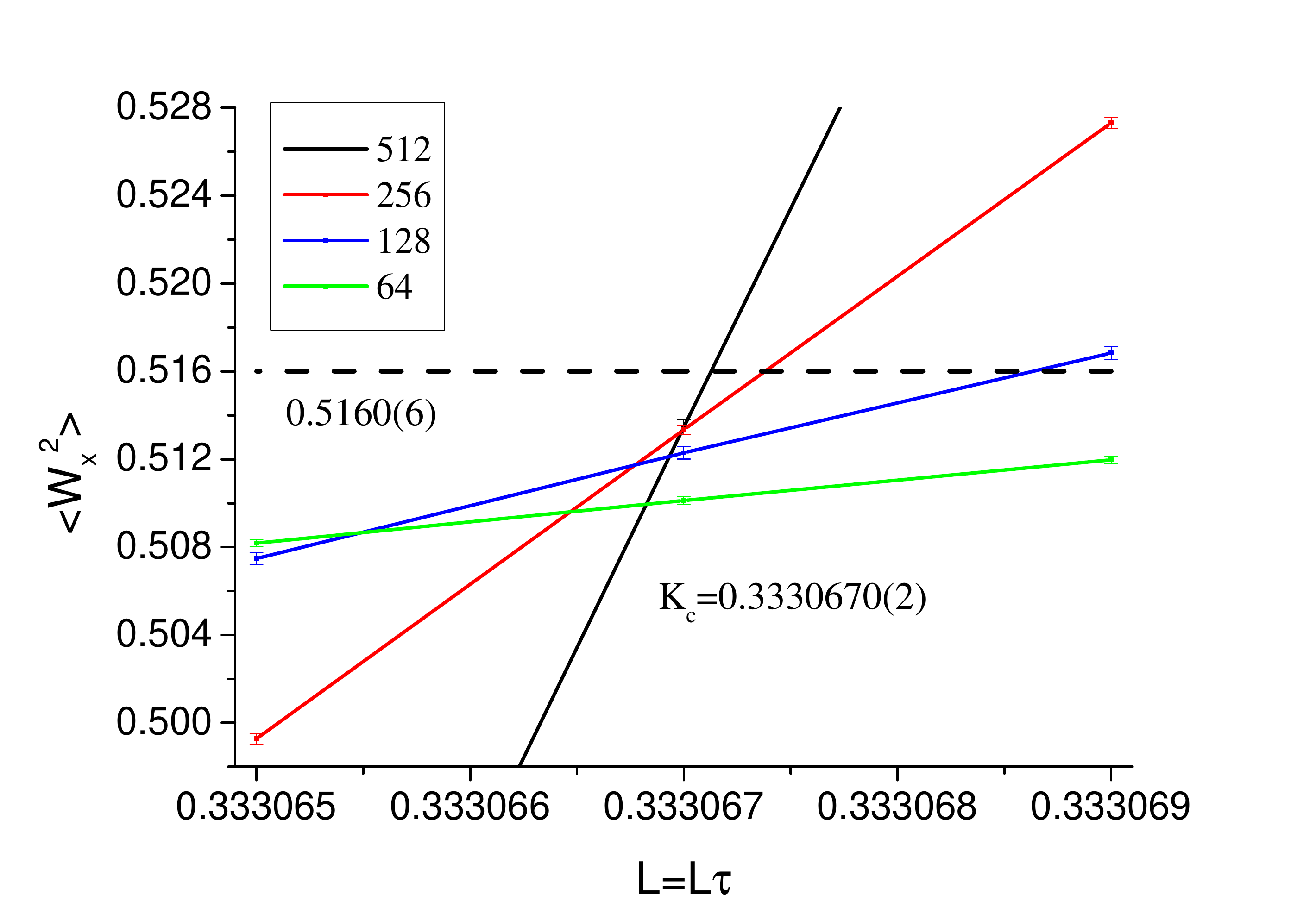}
\caption{\label{fig:1}(Color online) Statistics of winding numbers squared for different system sizes $L=L_{\tau}$. Fitting
the data using $a+b(K-K_c)L^{1/\nu}+cL^{-\omega}$ with $\nu=0.6717(3)$ and $a=0.5160(6)$\cite{evgeni} leads to
the critical point $K_c=0.3330670(2)$ of the J-current model [Eq. (\ref{jcurrent})].}
\end{figure}

{\it Systems -- } Our simulations were performed for the three-dimensional classical J-current model
with $L^2\times L_{\tau}$ sites,
\begin{equation}
H=\frac{1}{2K} \sum_{<ij>}^{\nabla {\mathbf J}=0} J^2_{<ij>} \;,
\label{jcurrent}
\end{equation}
where $J_{\langle ij \rangle} \in (-\infty, \infty)$ are integer valued bond currents between the
neighboring sites subject to the zero-divergence constraint such that the allowed configurations form closed loops. The same approach was used in Refs.~\cite{sorensen05,sorensen13}.
In Fig.~\ref{fig:1} we show our data for the winding numbers squared for different system sizes.
The crossing point determines the location of the critical parameter in the J-current model.
Our result $K_c=0.3330670(2)$ is consistent with previous estimates \cite{Alet,Neuhaus} but is far more accurate.
As explained in the supplementary material\cite{suppl}, if $K_c=0.33305$ is used instead, the data are dramatically
affected by that.

We also simulated the 2d quantum Bose-Hubbard model,
\begin{equation}\label{eq:BH}
H = - t \sum_{<ij>}  b_i^\dag b_j^{\,}
+ \frac{U}{2} \sum_i n_i(n_i-1) - \mu \sum_i n_i  \, ,
\end{equation}
where $-t$ is the hopping matrix element, $U$ the strength of the on-site
interaction, and $\mu$ the chemical potential. Its
quantum critical point at filling factor $\langle n \rangle =1$ is located at $U_c/t=16.7424(1)$, $\mu_c/t=6.21(2)$ \cite{Soyler1,Soyler2}.

{\it Methodology -- } For both classical and quantum models, the conductivity in Matsubara representation is given by~\cite{white,fisher91},
\begin{equation}
\label{eq:cond}
\sigma(i\omega_n)=2\pi\sigma_Q\frac{\langle -k_x\rangle-\Lambda_{xx}(i \omega _n)}{\omega_n} ,
\end{equation}
which is a direct consequence of the well-known Kubo formula. Matsubara frequencies are defined as
$\omega_n=2\pi n/\beta$ and $\omega_n=2\pi n/L_\tau$ for the bosonic and classical systems, respectively. $\langle k_x \rangle$ is the kinetic energy associated with a $x$-oriented bond, $\Lambda_{xx}(i \omega _n)$ is the fourier transform of imaginary time current-current correlation function~\cite{white}. The numerator has an unbiased estimator in the path
integral representation of the Bose-Hubbard model,
$\frac{1}{\beta L^2}\langle \vert \sum_k sgn(k)e^{i\omega_n \tau_k} \vert^2 \rangle$,
where the sum runs over all hopping transitions in a given configuration and $sgn(k)$ takes values
$+1$ or $-1$ depending on the positive or negative direction of the $k$-th transition.
A similar estimator can be applied to the J-current model. The conductivity on the real
frequency axis, $\sigma(\omega)$, requires an ill-conditioned analytical continuation
$i\omega_n \rightarrow \omega+i0^+$.

To obtain the universal conductivity in the quantum critical region, one has
to carefully extrapolate data to the infinite system size and zero temperature limits such that $L \to \infty$ is taken first, and
$L_{\tau} \to \infty$ next\cite{suppl}. The largest system size used in this study was $L=1024, L_{\tau}=512$ for the  J-current model and $L=400, \beta=20/t$ for the Bose-Hubbard model. We refer the reader to the supplementary material\cite{suppl} where
we describe the details of protocols used to eliminate finite-size and finite-temperature effects.

\begin{figure}[htbp]
\includegraphics[scale=0.5,angle=0,width=1.0\columnwidth]{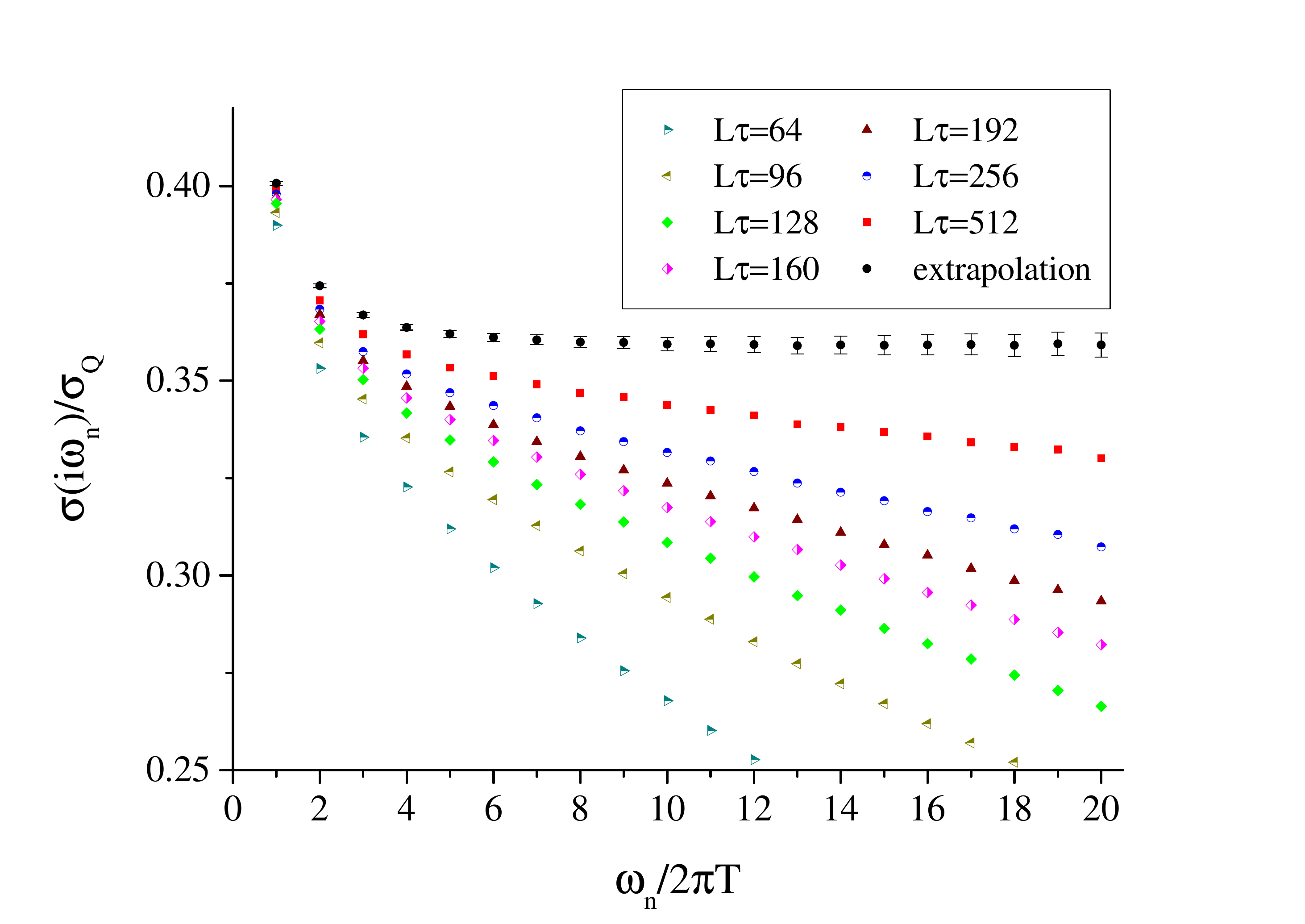}
\caption{\label{fig:2}(Color online) Thermodynamic limit data for the conductivity in Matsubara representation for increasing values of $L_{\tau}$), bottom to top.
The universal result is obtained by extrapolating the data to
the $L_{\tau}\to \infty$ limit (black filled circles). }
\end{figure}

{\it Jcurrent model --} Results for the quantum-critical conductivity in the Matsubara  representation
$\sigma(i\omega_n)$ are shown in Fig.~\ref{fig:2} for various values of $L_{\tau}$
along with the extrapolated zero-temperature limit.  The universal conductivity
quickly saturates to a plateau $\sigma(\infty)=0.359(4)$ for $\omega_n/2\pi T>6$,
as is expected from  Eq.~(\ref{eq:scaling}).

\begin{figure}[htbp]
\includegraphics[scale=0.5,angle=0,width=1.0\columnwidth]{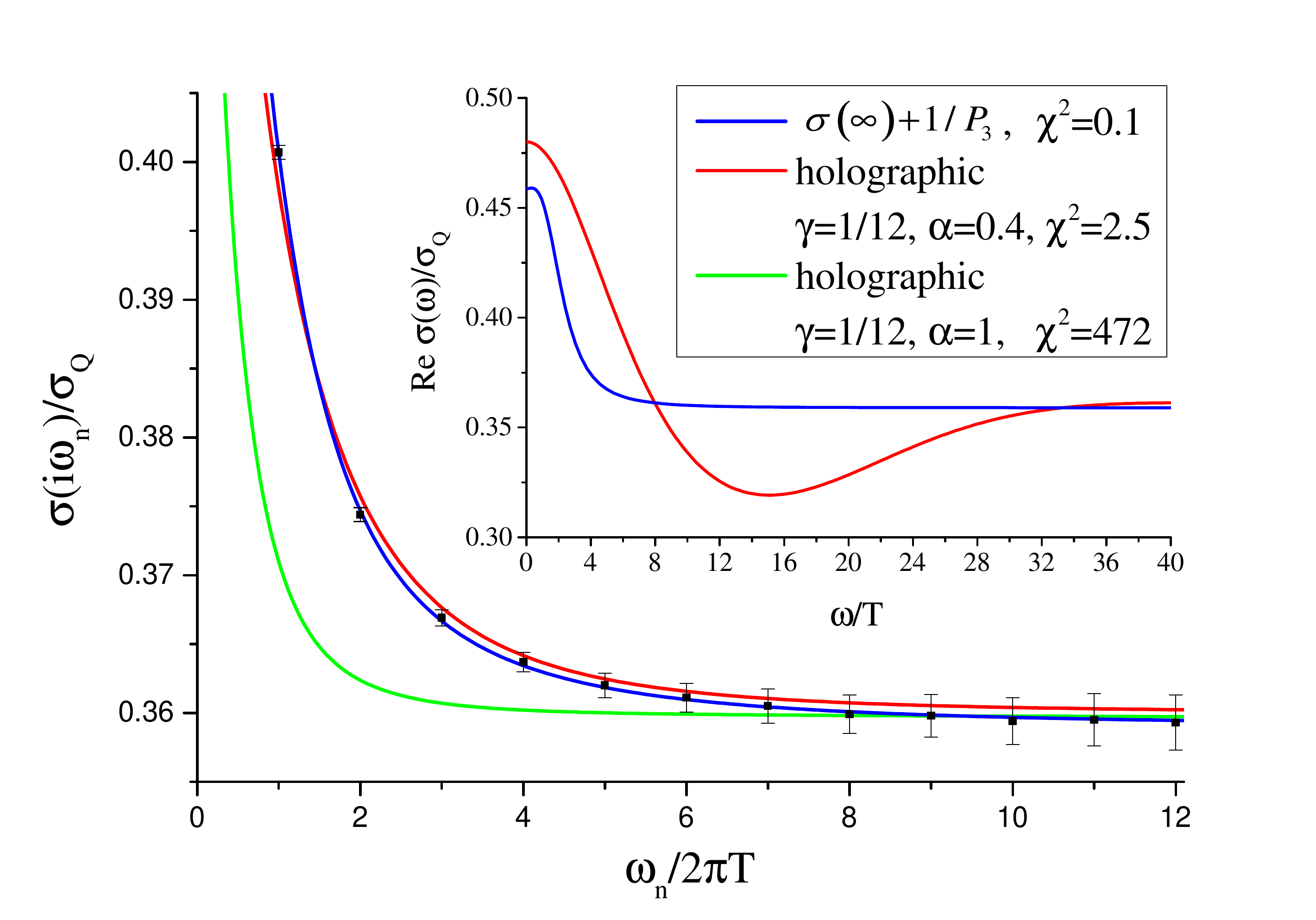}
\caption{\label{fig:3}(Color online) Comparison between the numerical result for the universal
conductivity in Matsubara space and holographic conductivity evaluated at complex frequencies.
We consider only the largest possible value of the holographic parameter $\gamma =1/12$ because it
offers the best fit. The holographic curve can be made to fit the data much better
if the temperature is scaled by a factor of $2.5$. We also show the fit based on the 3rd order polynomial described in the text. The inset presents the possible universal
real frequency conductivity curves obtained by analytical continuation of fitting functions.}
\end{figure}

\begin{figure}[htbp]
\includegraphics[scale=0.5,angle=0,width=1.0\columnwidth]{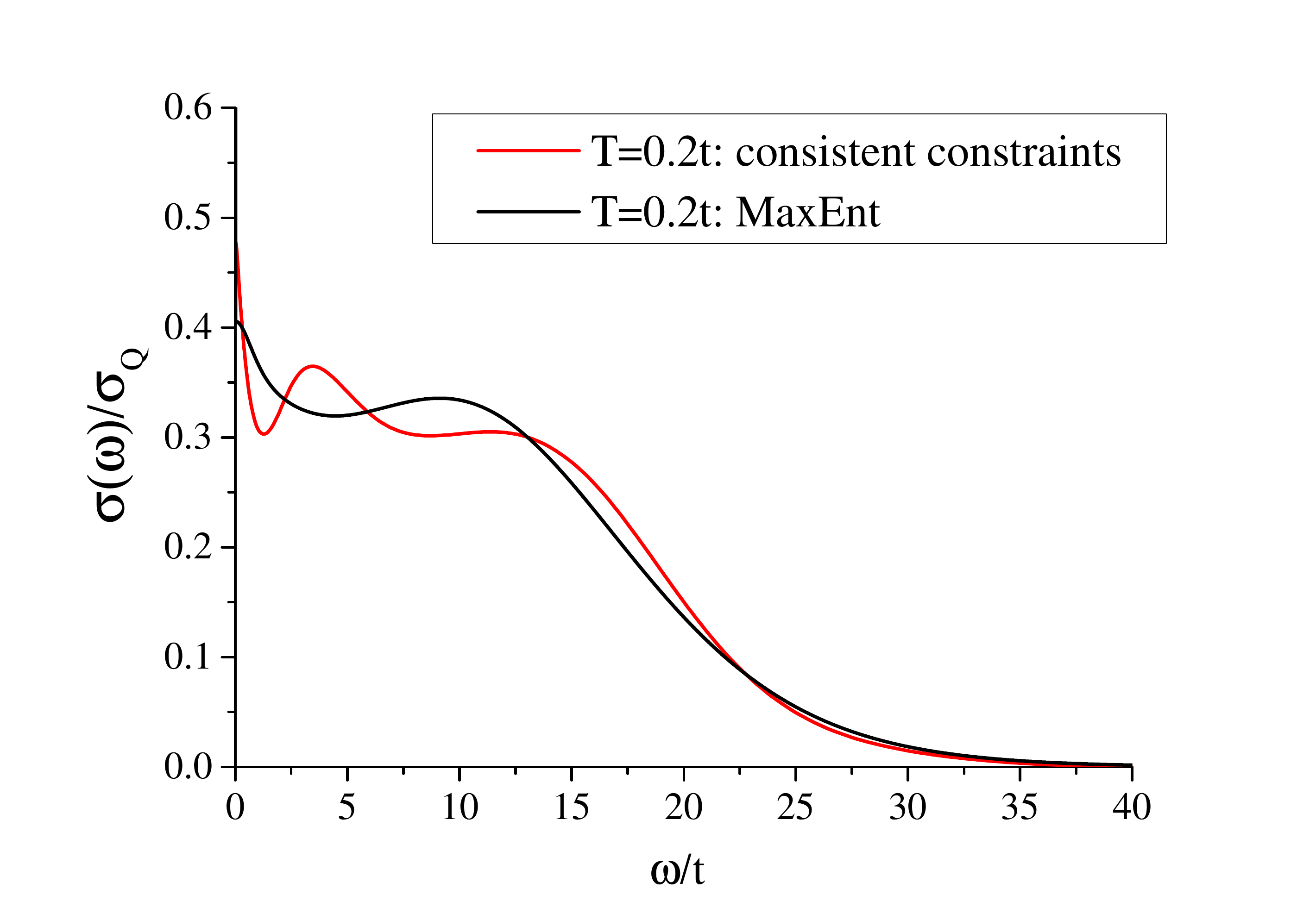}
\caption{\label{fig:5} Real frequency conductivity of the Bose Hubbard system at $T/t = 0.2$ and $L =100$. Both MaxEnt and consistent constraints analytical continuation methods~\cite{nikolay13} are tested.}
\end{figure}

In Fig.~\ref{fig:3} we show the comparison between the universal data for $\sigma(i\omega_n)$
and predictions based on the holographic gauge/gravity duality.
Clearly, the $T_B=T$ case is ruled out by the data.
However, if the temperature scale for dynamics problems is renormalized relative to the thermodynamic temperature
$T_B=T/\alpha$, as suggested in Ref.~\cite{sorensen13},
then the theory contains an additional freedom for modifying the shape of the holographic
curve by effectively rescaling the frequency axis by $\alpha$. For $\gamma=1/12$ we find that $\alpha$ is close to $0.4$.
Unfortunately, this additional fitting parameter puts the test to the limit:
given the benign shape of the imaginary frequency signal, much smaller error bars are required for furnishing a
conclusive test---this is a well-known problem in the analytic continuation procedure when
radically different functions in the real-frequency domain have very close shapes in the Matsubara
domain. For example, a simple analytic expression,
$\sigma(i\omega_n)=\sigma(\infty)+1/P_3(\omega_n/T)$, where $P_3(x)$ is the 3rd order polynomial,
perfectly fits all our data, has all poles in the lower half-plane, but results in a dramatically
different real-frequency signal, see inset in Fig.~\ref{fig:3}. Both outcomes suggest that transport is dominated by particle-like excitations.
When we attempt other analytic continuation procedures such as the one used
in the Higgs mode study~\cite{nikolay13}, we find that the result is rather unstable, see Fig.~\ref{fig:5}. We thus conclude that the shape of the real frequency
curve remains uncertain.

\begin{figure}[htbp]
\includegraphics[scale=0.5,angle=0,width=1.0\columnwidth]{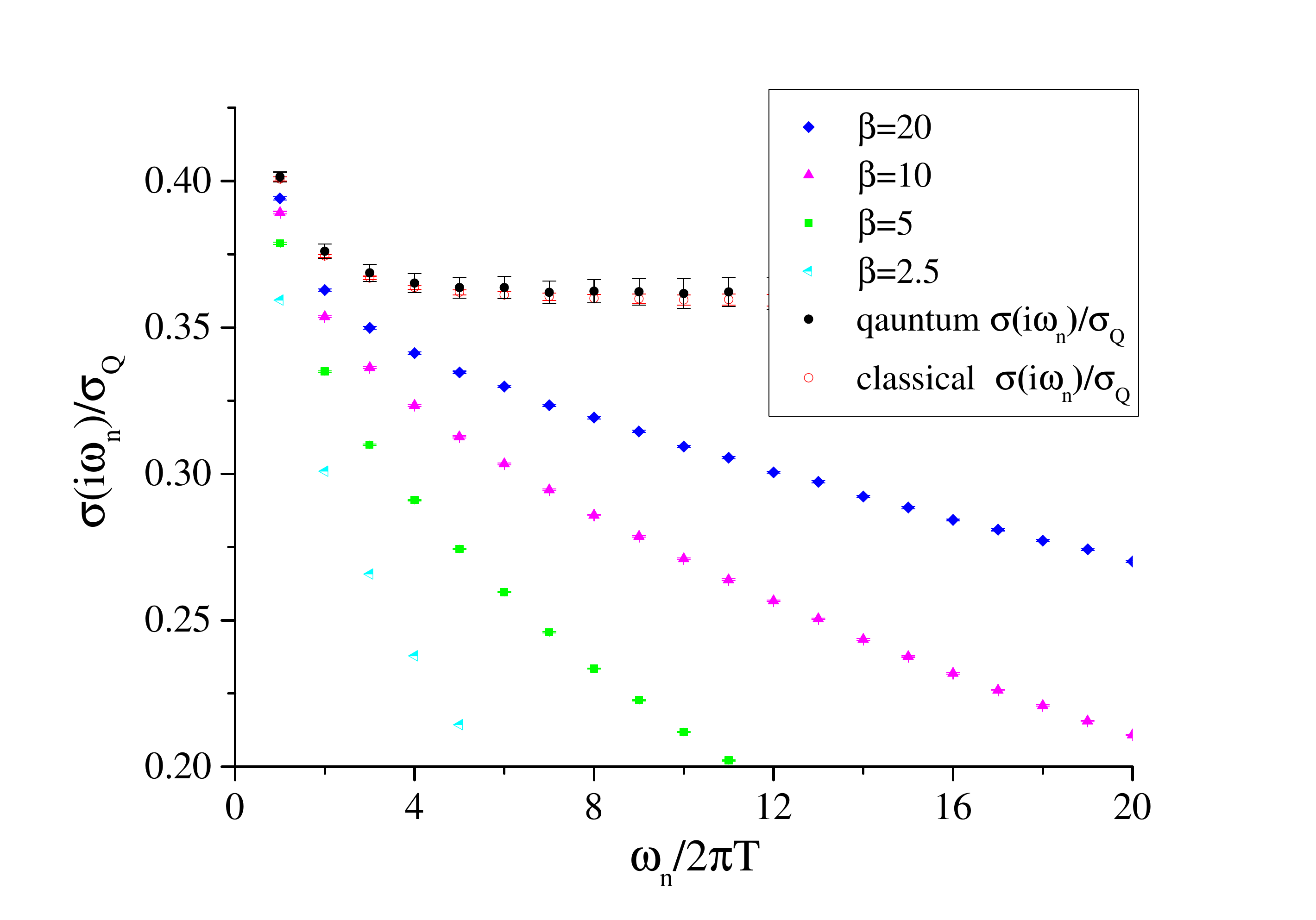}
\caption{\label{fig:4}(Color online)  Universal result for the conductivity in Matsubara representation from quantum Monte Carlo simulations for the Bose-Hubbard model. Temperature decreases from the bottom to the top, and the extrapolation to zero temperature is shown by black squares.}
\end{figure}

{\it Quantum model --} Simulations of the quantum Bose-Hubbard model are far more demanding numerically resulting in larger error bars.
Nevertheless, one can obtain reliable data for temperatures as low as $\beta=20/t$
in system sizes $L=400$ which are required for eliminating finite-size effects. The result of
extrapolation of the quantum model to the universal limit is shown in Fig.~\ref{fig:4}~\cite{suppl}.
As expected, the same result emerges from the quantum simulation, within error bars.

Unlike the classical model with discrete imaginary time direction,
$\sigma(i\omega_n)$ for a quantum system is physically meaningful for all Matsubara frequencies
and contains valuable information detailing the boundary between the universal and non-universal
parts of the signal. We employ the formula
\begin{equation}
\label{eq:ac}
\sigma(i\omega_n)=\frac{2}{\pi}\int_0^{\infty}\frac{\omega_n}{\omega^2+\omega_n^2}Re\sigma(\omega)d\omega   ,
\end{equation}
to obtain the real part of conductivity on the real frequency axis, $\sigma (\omega )$.
In Fig.~\ref{fig:5} we show $\sigma (\omega )$ for the Bose-Hubbard model at low-temperature $T=0.2t$
(for system size $L=100$ and critical point parameters). At frequencies $\omega > 2\pi T \approx 1.3t$
the analytical continuation procedure gets more stable and predicts, surprisingly,
that the plateau at $0.35(5)$ extends up to $\omega \approx 15t$.

{\it Experimental verification --} There are prospects that temperatures are within reach of experiments with ultracold atoms in an optical lattice once new cooling schemes are implemented
(this is a pressing issue for all proposals aimed at studies of strongly correlated states).
One idea of measuring the optical conductivity is based on shaking the center of magnetic trap horizontally to create an effective "AC electric field" and particle currents. The energy absorbed by the system (it can be deduced by measuring the temperature increase) is proportional to $Re\sigma(\omega)$. An amplitude modulation of the lattice laser in combination with a similar energy absorption protocol has successfully been applied to study the Higgs amplitude mode in the Bose-Hubbard model\cite{bloch,lode12}. Alternatively, phase modulation was also proposed in Ref.~\cite{findit} through the modulation of the position of a mirror reflecting the lattice laser beam, which allows us to measure $\omega^2 Re \sigma(\omega)$. However, the low frequency signal may be masked by the prefactor $\omega^2$ if the experiment is set up this way.

{\it Conlusions --} In conclusion, we have constructed the universal conductivity in Matsubara representation by carefully extrapolating finite system size data to the thermodynamic limit $L \rightarrow \infty$ and then taking the
$T \rightarrow 0$ limit.
Our result $\sigma(\infty)=0.359(4)\sigma_Q$ is the most accurate estimate of this quantity to date.
The shape of the universal conductivity function in Matsubara representation was measured with
reliable error bars for the first time and used to test the holographic theory\cite{sachdev11}.
The result of this comparison confirms that the theory has to include the possibility that the black brane
temperature $T_B$ is renormlaized relative to the thermodynamic temperature.
For $T_B\approx 2.5 T$ the holographic fit can be made marginally compatible with our data.
An unambiguous proof of validity, which requires, however, more accurate data or models with weaker finite size effects, would constitute a major breakthrough in establishing the analytic continuation procedure for transport properties at quantum critical points \cite{sorensen13}.
On existing data, simple analytic expressions work as well.
Finally, we also determine under what conditions $\sigma(\infty)$ can be measured with
ultracold atoms in optical lattices.

We wish to thank S. Sachdev, W. Witczak-Krempa, E. S. S{\o}rensen, B. V. Svistunov, Y. Huang, D. Schimmel, M. Endres and A. Singh for valuable discussions. Use was made of the MaxEnt application~\cite{maxent} in the ALPS libraries~\cite{alps2}. This work was supported in part by the National Science Foundation under Grant No. PHY-1314735,  FP7/Marie-Curie Grant No. 321918 (``FDIAGMC"), FP7/ERC Starting Grant No. 306897 (``QUSIMGAS"), the excellence cluster Nano-Initiative Munich (NIM), a grant from the Army Research Office with funding from the DARPA, NNSFC Grant No. 11275185, CAS, NKBRSFC Grant No. 2011CB921300 and AFOSR/DoD MURI ``Advanced Quantum Materials: A New Frontier for Ultracold Atoms" program. We also thank the hospitality of the Aspen Center for Physics (NSF Grant No. 1066293).

{\it Note added--} In the final stages of this work the independent results of Ref~\cite{sorensen13}
became publicly available. We thank the authors for sharing their results prior to publication with us.

\vspace*{2cm}

\section{\large{Supplementary material}}

This supplementary material contains technical details of two protocols: the first one is
used to obtain data for conductivity $\sigma (i\omega_n)$ in the Matsubara frequency representation in the thermodynamic limit $L\to \infty $ at fixed $L_{\tau}$ (in the classical case)
or fixed $T$ (in the quantum case), and the second one is used to extrapolate the thus obtained thermodynamic limit data to the universal $L_{\tau} \to \infty$ (zero temperature) limit. We also demonstrate that the
numerical data in Ref.~\cite{sorensen13} cannot be used for meaningfully testing the holographic theory because
the data is taken too deep on the insulating side of the transition for the available system sizes. The fact that the limits $L \to \infty$ and $L_{\tau} \to \infty$ have to be taken in this order requires a tremendous precision on the location of the critical point, which was underappreciated.

\begin{figure}[htbp]
\includegraphics[scale=0.5,angle=0,width=1.0\columnwidth]{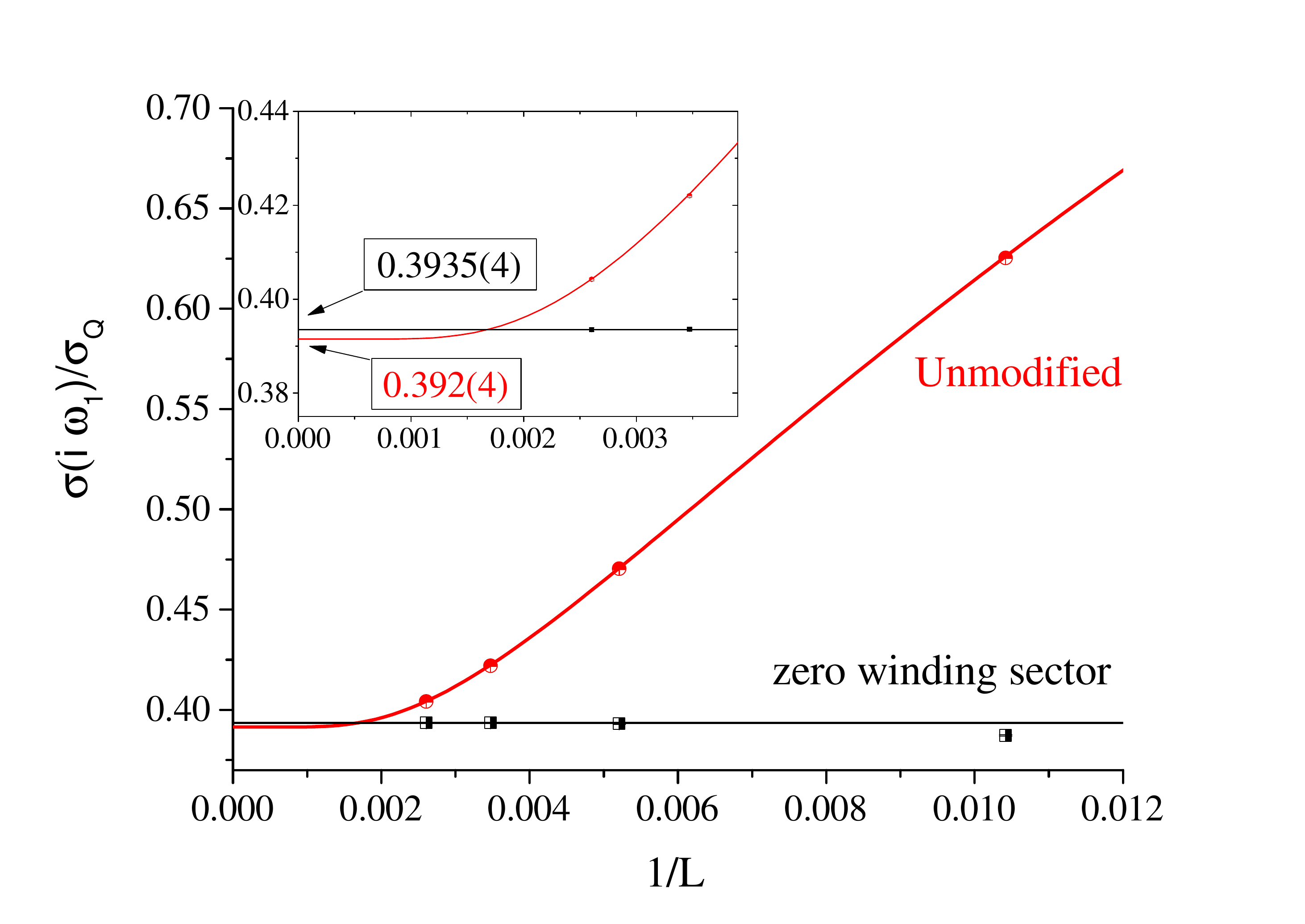}
\caption{\label{fig:1suppl}Extrapolating conductivity obtained for the first non-zero Matsubara frequency
$\omega_1$  to the thermodynamic limit $L/L_{\tau} \to \infty$ for a fixed system size $L_{\tau}=96$ in the $\tau$ direction (imaginary time). Unmodified [zero-winding sector] data are
plotted with red circles [black squares]. The result of the extrapolation using an exponential form with
$b=130$ is shown by the (red) solid line. In the zero-winding number sector the data saturate much faster
to the thermodynamic limit as is shown by the (black) horizontal line.}
\end{figure}

\section{Finite system size extrapolation}
\begin{figure}[htbp]
\includegraphics[scale=0.4,angle=0,width=0.8\columnwidth]{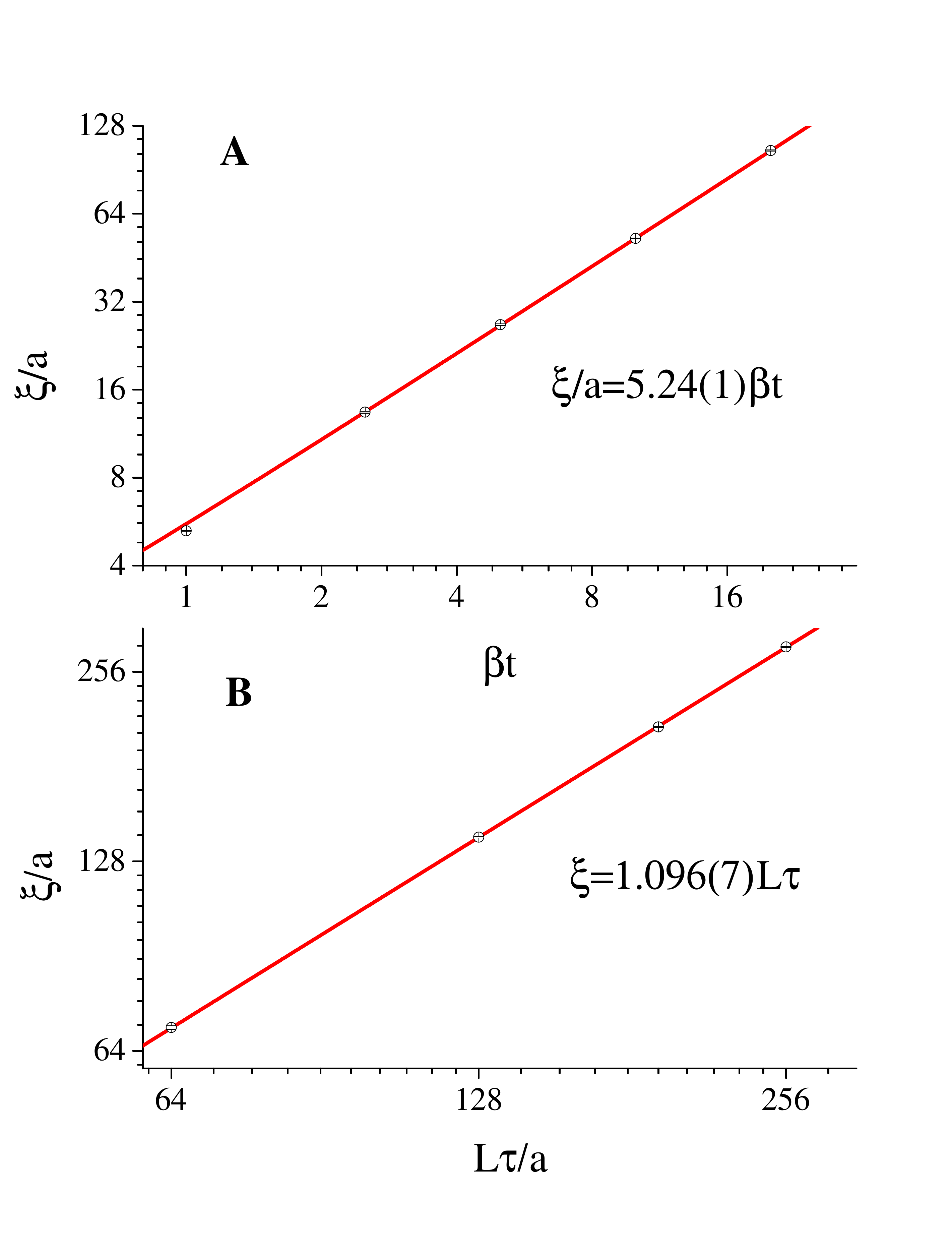}
\caption{\label{fig:2suppl} The dependence of the correlation length on temperature at the critical point of (A) the two-dimensional Bose-Hubbard model and (B) the (2+1)-dimensional J-current model. As expected, at low temperature
the dependence on $\beta= 1/T$ or $L_\tau$ is linear (solid line). Error bars are shown but are smaller than the
symbol sizes.}
\end{figure}

The extrapolation to the thermodynamic limit $L\to \infty$ for fixed $L_{\tau}$ can be done
in two ways. One way is to consider unmodified data for ever increasing values
of $L/L_{\tau}$ and to extrapolate them using an exponential scaling law, $e^{-L/b}/\sqrt{L}$,
\cite{Privman} with a temperature-dependent fitting parameter $b$.
In the quantum critical region the correlation length is directly proportional to
$L_{\tau}$ (or inverse temperature), and direct measurements of the correlation length
using the exponential decay of the single-particle density matrix indeed find that $\xi \approx 1.096(7) L_{\tau}$, see
Fig.~\ref{fig:2suppl}.
It is thus expected that $b$ also scales with $L_{\tau}$ and the data are best fit with
$b\approx 1.3 L_{\tau}$. A typical example of the exponential extrapolation
for the first non-zero Matsubara frequency at $L_{\tau}=96$ is shown in Fig.~\ref{fig:1suppl}.
The other way is to collect statistics for correlation functions only in the zero winding number sector.
It turns out that the second protocol is far more efficient in terms of system sizes required for reaching
the asymptotic thermodynamic values, see Fig.~\ref{fig:1suppl}. In the zero-winding sector
the data are essentially in the thermodynamic limit already for $L/L_{\tau}=2$, while unmodified data
extrapolate to the same result within error bars only if the point $L/L_{\tau}=4$ is included in the fit.
We checked for consistency between the two protocols for a number of points but for the largest values
of $L_{\tau}$ the simulation was done only in the zero-winding sector using $L/L_{\tau}=2$.

Quantum data were analyzed similarly.
First, we calibrate
the dependence of the correlation length $\xi$ on temperature at the critical point $(U/t)_c=16.7424$ \cite{Soyler2}. It is
deduced from the exponential decay of the single-particle density matrix at large distances and is expected
to increase $\propto 1/T$. The result of a linear fit shown in Fig.~\ref{fig:2suppl} leads to the
asymptotic dependence $\xi(T) \approx 5.24(1) \beta t$ in units of the lattice constant $a$.
The ratio $\xi /c\beta$, where $c$ is the sound velocity, is expected to be a universal
characteristic of the quantum critical point, the same as $\xi/L_{\tau}$ in the classical
J-current model. Using the value $c/a=4.8(2)t$ determined in Ref.~\cite{Soyler1}
we find that $\xi /c\beta =1.09(4)$, in perfect agreement with the classical result for $\xi/L_{\tau}$.
Given the exponential  convergence of $\sigma (i\omega_n)$ on $L$, we consider system sizes $L \approx 4 \xi$
(an equivalent of $L/L_{\tau} \approx 4.2$ for the classical system) for collecting unmodified data, and
system sizes $L \approx 2 \xi$ for  collecting data in the zero-winding sector.

\begin{figure}[htbp]
\includegraphics[scale=0.5,angle=0,width=1.0\columnwidth]{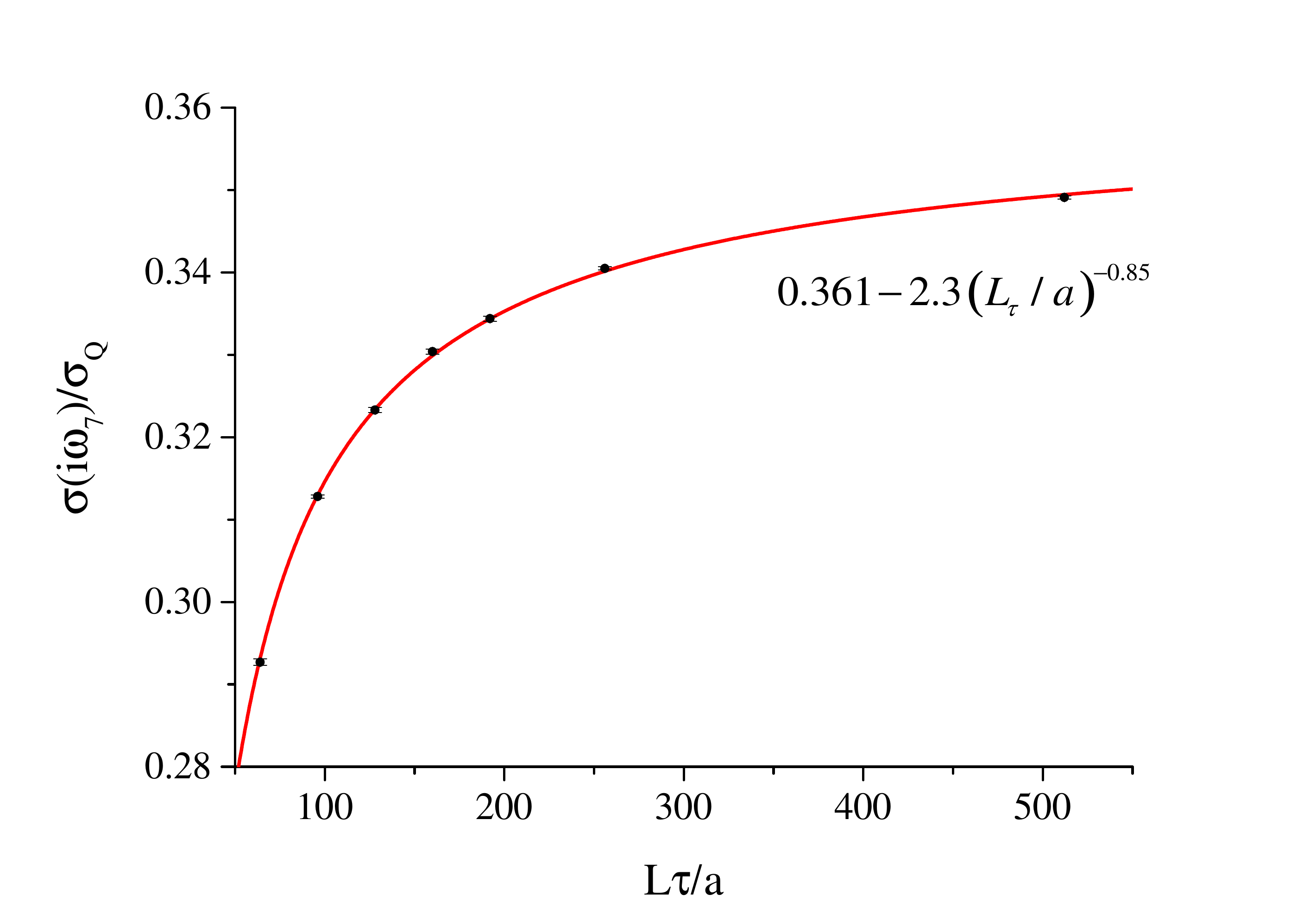}
\caption{\label{fig:3suppl}Extrapolating the conductivity in Matsubara representation to the universal
zero-temperature limit for fixed $\omega_n/2\pi T = 7$.}
\end{figure}
\begin{figure}[htbp]
\includegraphics[scale=0.5,angle=0,width=1.0\columnwidth]{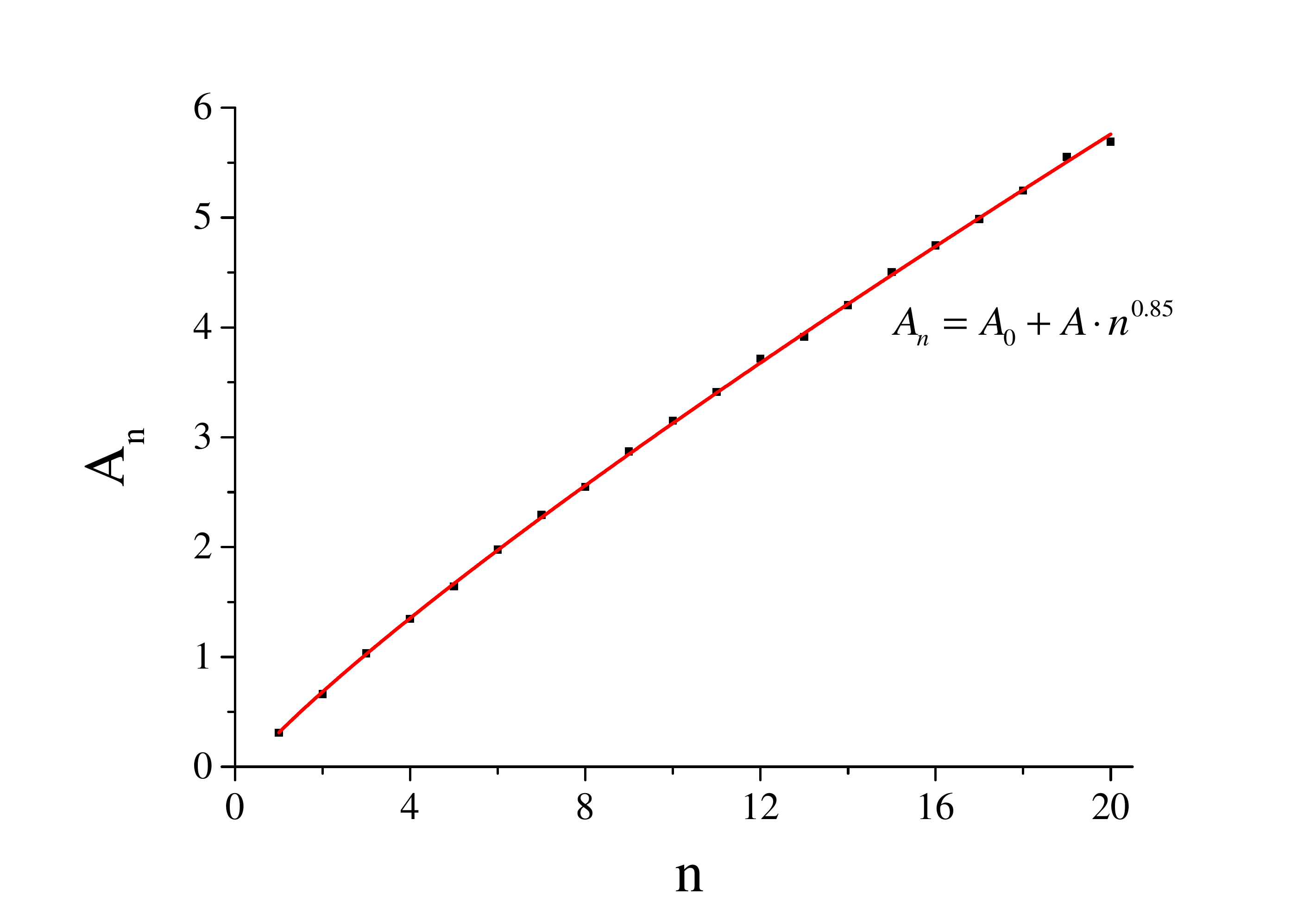}
\caption{\label{fig:3asuppl} $A_n$ amplitudes deduced by fitting each Matsubara
harmonic independently, see Fig.~\ref{fig:3suppl}. }
\end{figure}

\section{Zero temperature extrapolation}
Thermodynamic limit results for $\sigma(i\omega_n)$ have a smooth monotonic dependence
on $L_{\tau}$ and can thus reliably be
extrapolated to the universal $L_{\tau} \to \infty $ limit. We employ the standard
picture of critical phenomena which attributes finite $L_{\tau}$
corrections to the leading irrelevant scaling field for the three-dimensional $U(1)$
universality class; its exponent $\omega $ was calculated and measured in a number of studies and
estimates cluster around the field theoretical value $\omega \approx 0.80(2)$ \cite{Guida}.
In our analysis, we keep this exponent fixed at $\omega =0.85$ since with this choice we
observe that all fits are consistent with our error bars (if we include $\omega$ to the set of
fitting parameters we find that its value is in the range $0.90\pm 0.10$).
When we perform a two parameter fit,
$\sigma (i\omega_n/T, L_{\tau}) =  \sigma (i\omega_n/T, \infty) + A_n/L_{\tau}^{\omega}$, see Fig.~\ref{fig:3suppl},
we observe that the amplitudes $A_n$ form a smooth curve obeying the law $A_n=A_0+An^{\omega}$ with the same
exponent $\omega$, see Fig.~\ref{fig:3suppl},
suggesting that the leading correction is indeed dominated by the correlation volume $(L_{\tau}/n)^3$.
This allows us to perform a joint fit of all data points using $A_0, A$ and
the universal conductivity as the only adjustable parameters to suppress point-to-point fluctuations and
systematic error bars. The same protocol was applied to the quantum data.

\begin{figure}[htbp]
\includegraphics[scale=0.5,angle=0,width=1.0\columnwidth]{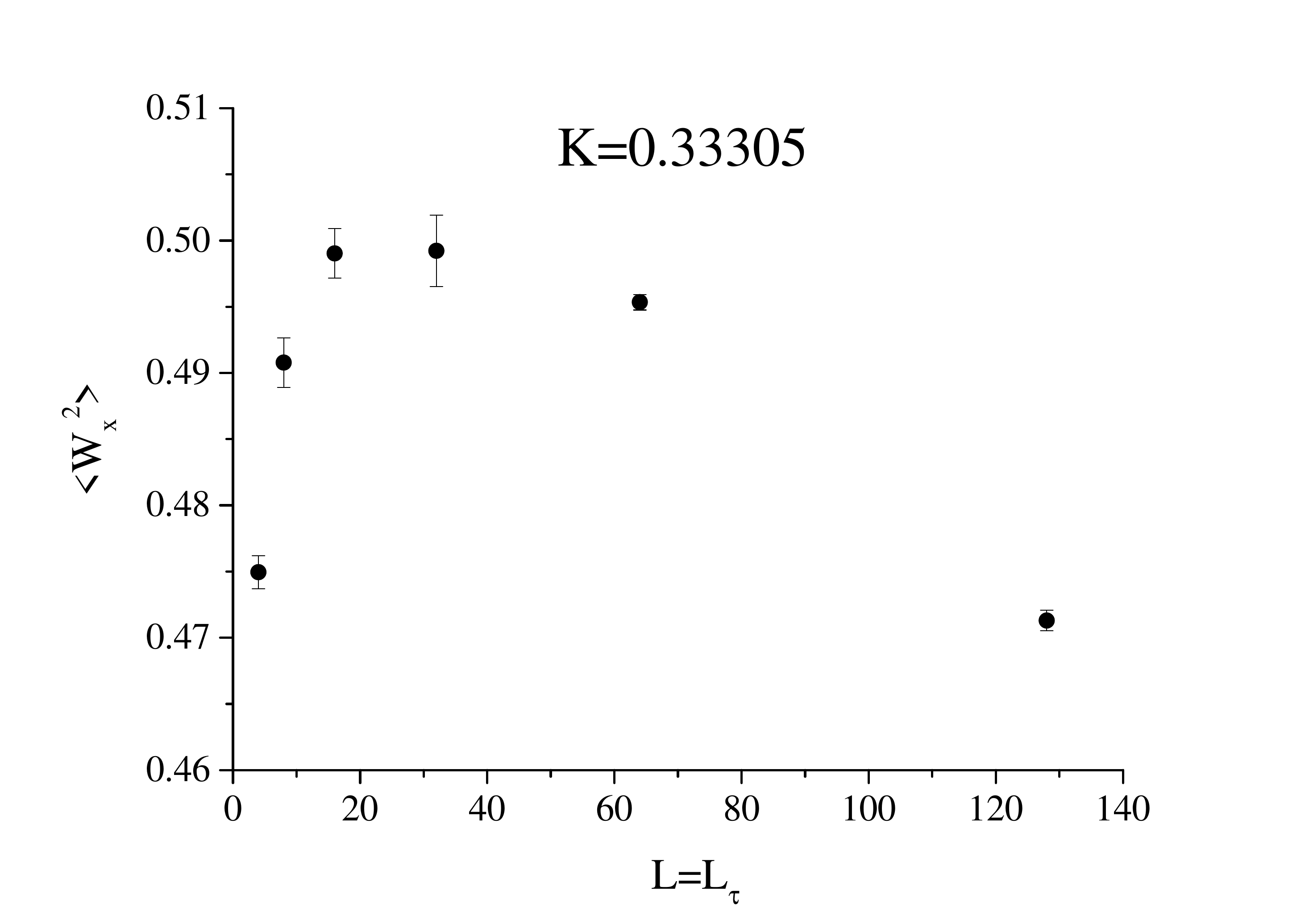}
\caption{\label{fig:4suppl} Statistics of winding numbers squared at $K=0.33305$ as a function
of system size. Instead of saturating to a constant, the data go through a maximum and start
decaying visibly for $L=128$, the largest system size of Ref.~\cite{sorensen13}. This is a clear signature of entering the insulating phase.}
\end{figure}
\begin{figure}[htbp]
\includegraphics[scale=0.5,angle=0,width=1.0\columnwidth]{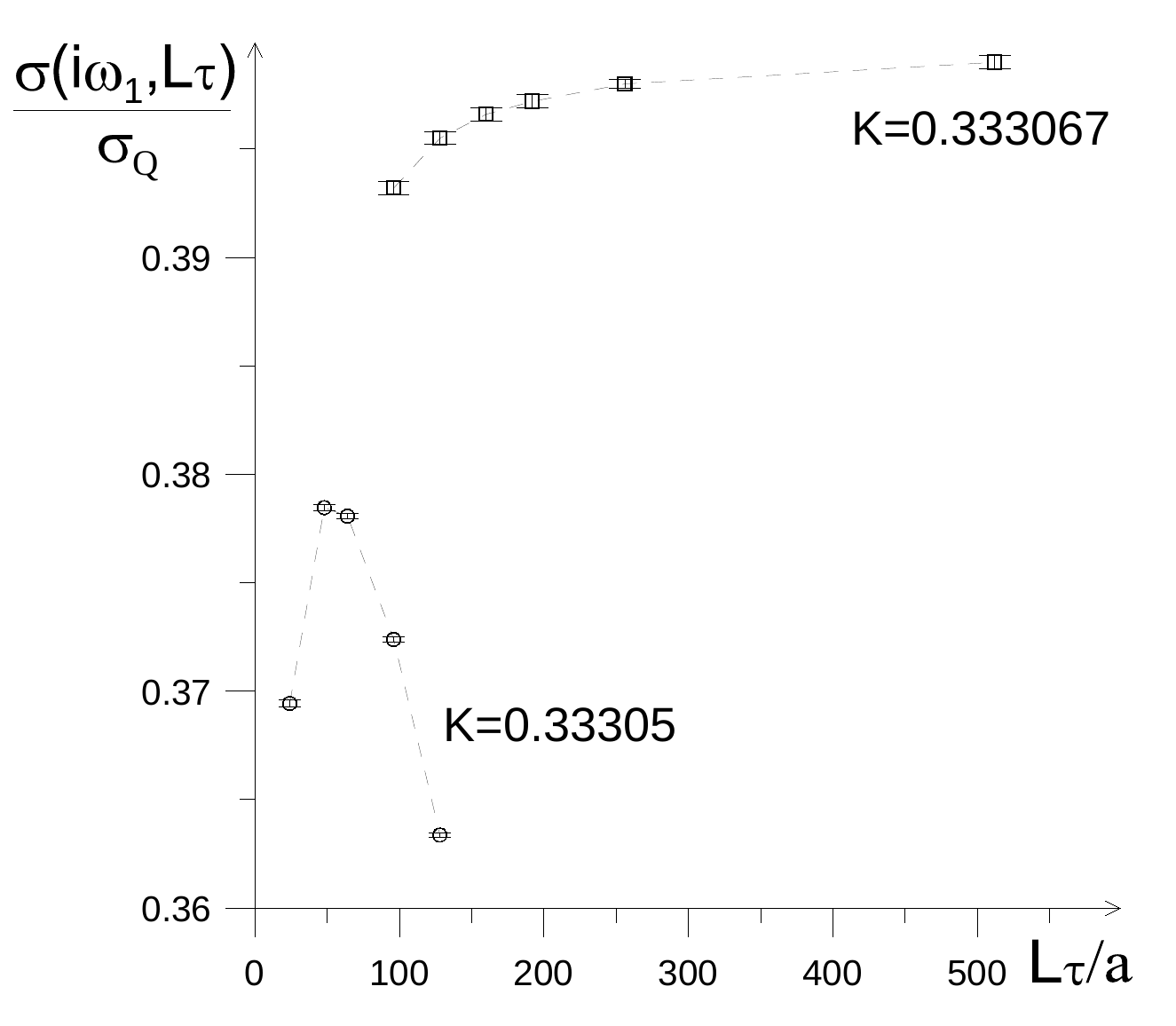}
\caption{\label{fig:4bsuppl} Side-by-side comparison of data for $\sigma(i\omega_1,L_{\tau})$
at $K=0.33305$ (lower curve) and $K=0.333067$ (upper curve). Clearly, the results for the lower curve cannot be extrapolated in a meaningful fashion unless the largest system sizes are excluded.}
\end{figure}
\begin{figure}[htbp]
\includegraphics[scale=0.5,angle=0,width=1.0\columnwidth]{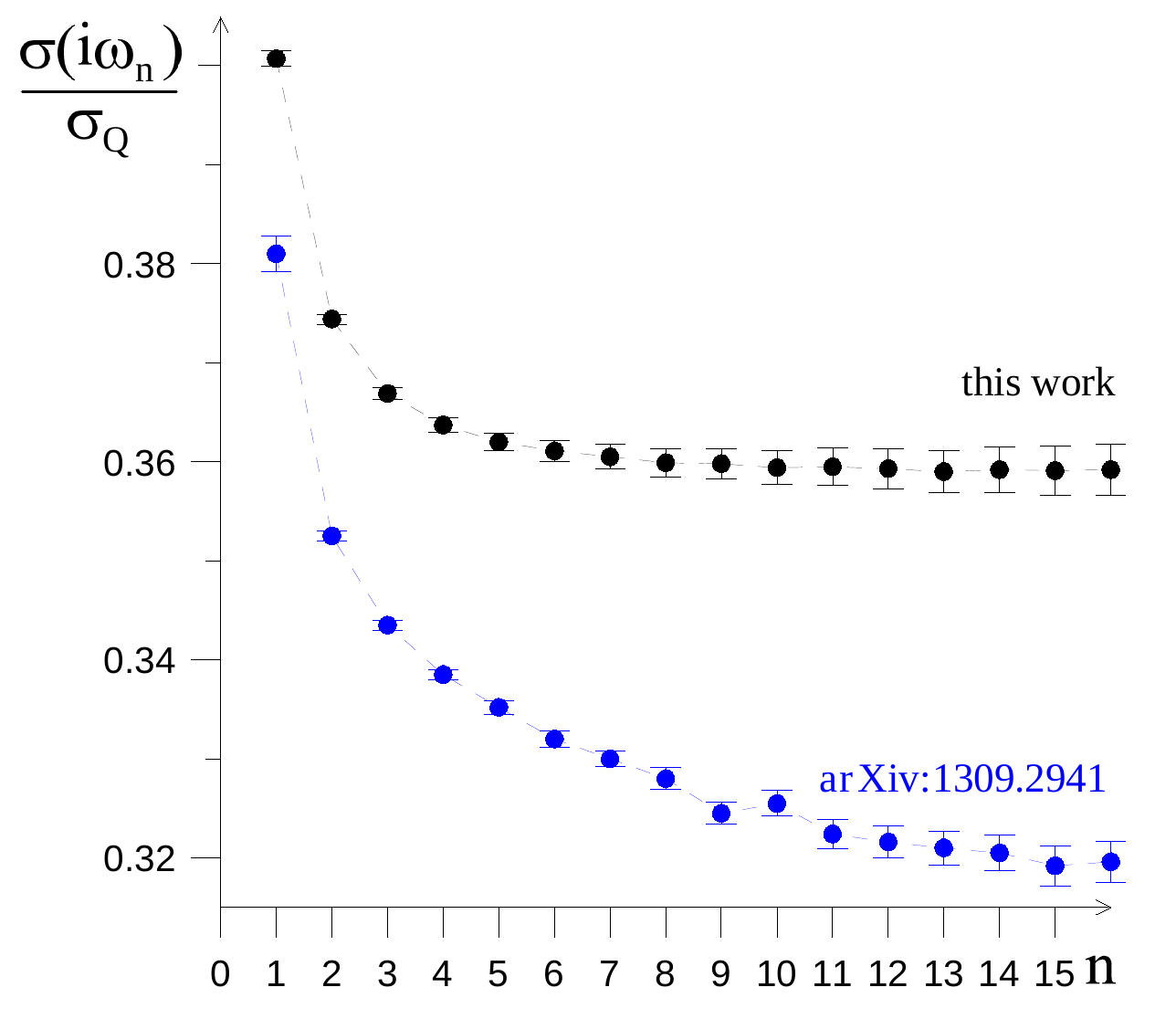}
\caption{\label{fig:5suppl} Direct comparison of the universal conductivity in Matsubara representation
computed in this work with results obtained in Ref.~\cite{sorensen13}. Clearly, the disagreement
is way beyond the claimed error bars, and, most importantly, amounts to a dramatically different
shape of the universal curve relative to the high-frequency plateau value. }
\end{figure}

\section{Discussion of the Results of Ref.~\cite{sorensen13}}
As explained in the main text, the critical point of the J-current model is located at
$K_c=0.3330670(2)$ while calculations performed in Ref.~\cite{sorensen13} were done at $K=0.33305$: {\it i.e.},
on the insulating side of the transition point. This might seem to be an irrelevant difference but it {\it does
affect} data for the universal conductivity to the extent that the shape of the universal $\sigma(i\omega_n)$
function is dramatically modified. As is clear from Fig.~\ref{fig:3suppl}, the finite $L_{\tau}$ dependence is strong
and large system sizes are required for a reliable extrapolation to the universal limit. When simulations are
done at $K=0.33305$, the results for winding numbers squared (which are characteristic of current fluctuations at the largest scales) and $\sigma(i\omega_n)$ for the lowest frequencies are significantly reduced relative to their critical behavior already for $L \approx 64$,
see Figs.~\ref{fig:4suppl} and \ref{fig:4bsuppl}. Lower values of
$\sigma(i\omega_n , L_{\tau})$ for large $L_{\tau}$ are then interpreted
as a sign of convergence to the universal limit in the fitting procedure of
Ref.~\cite{sorensen13}. As a result, the extrapolated values end up to be significantly lower than the correct ones. For the lowest frequencies, the difference is certainly not attributable to the extrapolation procedures alone because our data for $L_{\tau}=512$ are visibly above the extrapolated data of Ref.~\cite{sorensen13}. This explains why the data in Ref.~\cite{sorensen13} are displaced relative to the universal result way outside of their error bars. Being away from the critical
point is not an issue at high frequencies where the curve settles at the quantum critical plateau. In this region, the extrapolated data of Ref.~\cite{sorensen13} are lower than ours because Ref.~\cite{sorensen13} assumed that finite $L_{\tau}$ corrections are exponentially small in
$L_{\tau}$, which is impossible in the gapless quantum critical phase and contrary
to  the established picture of critical phenomena.

Summarizing, the entire shape of the universal function in Ref.~\cite{sorensen13} relative to the plateau value turns out to be
dramatically altered by systematic errors induced by performing simulations at $K=0.33305$ and
using incorrect extrapolation procedure,
see Fig.~\ref{fig:5suppl}. Even the most robust quantity $\sigma(\infty )$ differs by a surprisingly large 10\%, well outside of claimed error bars,  which can be attributed entirely to the location of the QCP (with a relative difference of only $10^{-5}$) and the sensitive extrapolations. When allowing for a  free parameter to rescale temperature in the functional form of the holographic theory this leads to a difference of $250\%$ between the two results.
Based on the numerical data reported in Ref.~\cite{sorensen13} a meaningful test of the holographic theory
simply cannot be done, and it comes hence as no surprise that the conclusions deduced from such a comparison in our work are dramatically different from those in Ref.~\cite{sorensen13}.


\begin{thebibliography}{99}

\bibitem{sachdev11}
R. C. Myers, S. Sachdev, and A. Singh, Phys. Rev. D {\bf 83}, 066017 (2011).

\bibitem{book}
S. Sachdev, {\it Quantum Phase Transitions} (Cambridge University Press, Cambridge, 2011), 2nd ed.

\bibitem{sachdev97}
K. Damle and S. Sachdev, Phys. Rev. B {\bf 56}, 8714 (1997)

\bibitem{fisher91}
M.C. Cha, M. P. A. Fisher, S. M. Girvin, M. Wallin, and A. P. Young, Phys. Rev. B {\bf 44}, 6883 (1991).

\bibitem{Fazio} R. Fazio and D. Zappala, Phys. Rev. {\bf B 53}, 8883 (1996).

\bibitem{william12}
W. Witczak-Krempa, P. Ghaemi, T. Senthil, and Y. B. Kim, Phys. Rev. B {\bf 86}, 245102 (2012)

\bibitem{mald}
J. M. Maldacena, Adv. Theor. Math. Phys. {\bf 2},231-252 (1998).

\bibitem{review1}
S. Sachdev, Annu. Rev. Condens Matter Phys. {\bf 3},9 (2012).

\bibitem{review2}
A. Adams, L. D. Carr, T. Schaefer, P. Steinberg and J. E. Thomas, New J. Phys. {\bf 14}, 115009 (2012).

\bibitem{engel}
L. W. Engel, D. Shahar, C. Kurdak, and D. C. Tsui, Phys. Rev. Lett. {\bf 71}, 2638 (1993).

\bibitem{crane}
R. Crane, N. P. Armitage, A. Johansson, G. Sambandamurthy, D. Shahar, and G. Gr{\"u}ner, Phys. Rev.
B {\bf 75}, 184530 (2007).

\bibitem{sorensen05}
J. \v{S}makov and E. S{\o}rensen, Phys. Rev. Lett. {\bf 95}, 180603 (2005).

\bibitem{sorensen13}
W. Witczak-Krempa, E. S{\o}rensen and S. Sachdev, arXiv:1309.2941 (2013)

\bibitem{suppl}
See Supplemental Material for the numerical analysis and a quantitative comparison with Ref.~\cite{sorensen13}.

\bibitem{Subir13} W. Witczak-Krempa and S. Sachdev,
Phys. Rev. {\bf B 87}, 155149 (2013).

\bibitem{batr} G.G. Batrouni, B. Larson, R.T. Scalettar, J. Tobochnik, and J. Wang
Phys. Rev. {\bf B 48}, 9628 (1993).

\bibitem{gazit} S. Gazit, D. Podolsky, A. Auerbach, and D.A. Arovas, Phys. Rev. {\bf B 88}, 235108 (2013).

\bibitem{Alet} F. Alet and E.S. S{\o}rensen, Phys. Rev. {\bf E 68}, 026702 (2003).

\bibitem{Neuhaus} T. Neuhaus, A. Rajantie, and K. Rummukainen,
 Phys. Rev. {\bf B 67}, 014525 (2003) .

\bibitem{Soyler1}
B. Capogrosso-Sansone, S.G. S\"{o}yler, N.V. Prokof'ev, and B.V. Svistunov,
Phys. Rev. {\bf A 77}, 015602 (2008).

\bibitem{Soyler2} S.G. S\"{o}yler, M. Kiselev, N.V. Prokof'ev, and B.V. Svistunov,
Phys. Rev. Lett. {\bf 107}, 185301, (2011).


\bibitem{evgeni}
E. Burovski, J. Machta, N.V. Prokof'ev,  and  B.V. Svistunov,
Phys. Rev. {\bf B 74} 132502 (2006).

\bibitem{white}
D.J. Scalapino, S. R. White, and S. C. Zhang, Phys. Rev. Lett. {\bf 68}, 2830 (1992); Phys. Rev. B {\bf 47}, 7995 (1993).

\bibitem{nikolay13}
N.V. Prokofiev, B.V. Svistunov, JETP Lett. {\bf 97}, 747 (2013); arXiv:1304.5198 (2013).

\bibitem{findit} A. Tokuno and T. Giamarchi,
Phys. Rev. Lett. {\bf 106}, 205301 (2011).

\bibitem{bloch}
M. Endres, T. Fukuhara, D. Pekker, M. Cheneau, P. Schau${\beta}$, C. Gross, E. Demler, S. Kuhr, and I. Bloch,
Nature {\bf 487}, 454-458 (2012).

\bibitem{lode12}
L. Pollet and N. Prokof'ev, Phys. Rev. Lett. {\bf 109}, 010401 (2012)

\bibitem{maxent}
M. Jarrell and J. E. Gubernatis,
Phys. Rep, {\bf 269}, 133 (1996).

\bibitem{alps2} B. Bauer {\it et al.},
J. Stat. Mech. , P05001 (2011).

\bibitem{Privman} {\it Finite Size Scaling and Numerical Simulations of Statistical
Systems}, ed. V. Privman, World Scientific, Singapore (1990).

\bibitem{Guida} R. Guida and J. Zinn-Justin, J. Phys. {\bf A 31}, 8103 (1998).

\end{thebibliography}
\end{document}